\documentclass[aps,pra,reprint,superscriptaddress]{revtex4-1}
\usepackage{url,amsmath,amssymb,bm}

\usepackage{multirow,graphicx,xspace}

\newcommand{\rrscan}{r$^2$SCAN\xspace}

\newcommand{\suxc}{_\text{xc}}
\newcommand{\sux}{_\text{x}}
\newcommand{\suc}{_\text{c}}
\newcommand{\sus}{_\text{s}}
\newcommand{\nup}{n_\uparrow}
\newcommand{\ndn}{n_\downarrow}
\newcommand{\br}{\bm{r}}
\newcommand{\bR}{\bm{R}}
\newcommand{\kf}{k_\mathrm{F}}
\newcommand{\rs}{r_\mathrm{s}}
\newcommand{\expec}[2]{\left\langle #1 \left| #2 \right| #1 \right\rangle}

\begin{document}

\title{Predictive Power of the Exact Constraints and Appropriate Norms in Density Functional Theory}

\author{Aaron D. Kaplan}
\email{kaplan@temple.edu}
\affiliation{Department of Physics, Temple University, Philadelphia, Pennsylvania 19122 USA}

\author{Mel Levy}
\affiliation{Department of Chemistry and Quantum Theory Group, Tulane University, New Orleans, Louisiana 70118 USA}

\author{John P. Perdew}
\affiliation{Department of Physics, Temple University, Philadelphia, Pennsylvania 19122 USA}
\affiliation{Department of Chemistry, Temple University, Philadelphia, Pennsylvania 19122 USA}

\begin{abstract}
  Ground-state Kohn-Sham density functional theory provides, in principle, the exact ground-state energy and electronic spin-densities of real interacting electrons in a static external potential.
  In practice, the exact density functional for the exchange-correlation (xc) energy must be approximated in a computationally efficient way.
  About twenty mathematical properties of the exact xc functional are known.
  In this work, we review and discuss these known constraints on the xc energy and hole.
  By analyzing a sequence of increasingly sophisticated density functional approximations (DFAs), we argue that: (1) the satisfaction of more exact constraints and appropriate norms makes a functional more predictive over the immense space of many-electron systems; (2) fitting to bonded systems yields an interpolative DFA that may not extrapolate well to systems unlike those in the fitting set.
  We discuss how the class of well-described systems has grown along with constraint satisfaction, and the possibilities for future functional development.
\end{abstract}

\maketitle

\onecolumngrid

\tableofcontents

\section{Introduction}

Atoms, molecules, and materials are collections of light-mass (quantum mechanical but typically non-relativistic) electrons and heavier-mass (almost classical) nuclei, interacting via Coulombic forces of attraction and repulsion.
Ignoring the nuclear motion, the Hamiltonian operator is often well represented by
\begin{equation}
  \hat{H} = \sum_{i=1}^N \left\{ -\frac{1}{2}\nabla_i^2 + v(\br_i)
    + \sum_{j > i}\frac{1}{|\br_i - \br_j|} \right\}
    + \sum_{\alpha=1}^M \sum_{\beta > \alpha}\frac{Z_\alpha Z_\beta}{|\bR_\alpha - \bR_\beta|}.
    \label{eq:full_hamil}
\end{equation}
The first term on the right-hand side of Eq. \ref{eq:full_hamil} is the kinetic energy of the electrons, labelled by indices $i=1,...,N$  at positions $\br_i$.
The second term is the interaction of the electrons with a scalar external potential $v(\br_i)$, which is usually just the Coulomb attraction to the nuclei, with $\alpha=1,...,M$ at positions $\bR_\alpha$ with charges $Z_\alpha$:
\begin{equation}
  v(\br) = -\sum_{\alpha=1}^M \frac{Z_\alpha}{|\br - \bR_\alpha|}.
\end{equation}
The third term of Eq. \ref{eq:full_hamil} is the Coulomb repulsion between electrons, and the last term is the Coulomb repulsion between the nuclei.
All equations are in Hartree atomic units.
In principle, all the properties of all the electron states of well-defined energy $E_k$ can be found by solving the many-electron Schr\"odinger eigenvalue problem
\begin{equation}
  \hat{H}\Psi_k(\br_1\sigma_1,...,\br_N\sigma_N) = E_k
    \Psi_k(\br_1\sigma_1,...,\br_N\sigma_N) \label{eq:se_eq}
\end{equation}
for the eigenvalues $E_k$ and eigenfunctions or wavefunctions $\Psi_k$, with the latter constrained to be antisymmetric under exchange of any two indices
\begin{equation}
  \Psi_k(\br_1\sigma_1,...,\br_i\sigma_i,...,\br_j\sigma_j,...,\br_N\sigma_N)
  = - \Psi_k(\br_1\sigma_1,...,\br_j\sigma_j,...,\br_i\sigma_i,...,\br_N\sigma_N).
  \label{eq:wvfn_exch}
\end{equation}
Here, $\sigma_i = \downarrow,\uparrow$ is the $z$-component of the spin of electron $i$.
A normalized wavefunction $\Psi$ yields an electronic spin-density
\begin{equation}
  n_\sigma(\br) = N \sum_{\sigma_2...\sigma_N} \int d\br_2 ...d\br_N \,
    |\Psi(\br\sigma,\br_2\sigma_2,...,\br_N\sigma_N)|^2, \label{eq:spin_dens}
\end{equation}
and the total electron density is $n(\br) = \nup(\br) + \ndn(\br)$, so that the average number of electrons in volume element $d\br$ is $n(\br)d\br$.
A density functional $E[n]$ is a rule that assigns a number $E$ (e.g., an energy) to a density function $n(\br)$.
For example, an integral over three-dimensional space of an integrand that can be found in closed form from the electron density is an \textit{explicit} functional thereof.

The correlated wavefunction theory \cite{szabo1982} of the previous paragraph promises ``the right answer for the right reason,'' with high accuracy, but at a computational cost that rises steeply with electron number due to the coupling of each electron to the other electrons.
The density functional theory (DFT) of Hohenberg, Kohn, and Sham \cite{hohenberg1964,kohn1965,parr1989,perdew2003,jones2015} and its generalizations, for the ground-state energy and electron density, starts from an ``exact'' DFT that is formally equivalent to Eqs. \ref{eq:full_hamil}, \ref{eq:se_eq}, and \ref{eq:wvfn_exch}, and then makes approximations intended to yield ``almost the right answer, for almost the right reason, at almost the right price, for almost all systems of interest.''
Improvements of approximations based upon the underlying exact theory have been moving slowly but relentlessly forward for over fifty years.
The utility of density functional theory is shown by the fact that it is perhaps the most widely cited area of physics, chemistry, and materials science \cite{noorden2014}.

The ground-state energy and density determine the most important properties of an atom, molecule, or material: the total energy yields energy differences including those to add or remove electrons or to break up a molecule or material into smaller fragments, such as separate atoms.
The gradient of the total energy with respect to a nuclear position, or equivalently the electrostatic force exerted on a nucleus by the electron density and by the other nuclei, yields the force on a classical nucleus, and hence the equilibrium shape of a molecule and the crystal structure of a solid, or the nuclear motion in a solid or liquid via \textit{ab initio} molecular dynamics.
The standard extension to spin-density functional theory \cite{barth1972} for a spin-dependent external potential $v_\sigma(\br)$ predicts many magnetic properties, and boosts the accuracy of the approximations even when the external potential is not spin-dependent.

In 1964, Hohenberg and Kohn \cite{hohenberg1964} proved two theorems for non-degenerate cases (later extended \cite{levy1979} to degenerate cases): (1) the external potential $v(\br)$ is determined, to within an arbitrary additive constant, by the ground-state electron density;
(2) the ground-state density and energy for external potential $v(\br)$ and electron number $N$ are found by minimizing
\begin{equation}
  E_v[n] = F[n] + \int d\br \, n(\br) v(\br) \label{eq:hk_etot}
\end{equation}
over $N$-electron densities $n(\br)$.
Here, $F[n]$ is a universal (independent of the external potential) functional for all non-degenerate electronic ground-state densities in the presence of external scalar potentials.
The Euler-Lagrange equation for this constrained minimization,
\begin{equation}
  \frac{\delta F}{\delta n(\br)} + v(\br) = \mu \label{eq:hk_el_eq}
\end{equation}
(where the chemical potential $\mu$ must be adjusted to fix the electron number $N$), illustrates theorem (1) above.
The functional derivative is defined by
\begin{equation}
  \delta F = \int d\br \left\{ \frac{\delta F}{\delta n(\br)} \right\} \delta n(\br).
\end{equation}
Equations \ref{eq:hk_etot} and \ref{eq:hk_el_eq} are not useful without a good prescription for $F[n]$.
In 1965, Kohn and Sham \cite{kohn1965} found a good prescription by positing a fictitious system of non-interacting electrons with the same ground-state density as the real, interacting system, but subject to an effective potential $v\sus([n],\br)$.
The notation used here means that $v\sus$ is a function of the position $\br$ and a functional of the electron density $n(\br)$.
They wrote
\begin{equation}
  F[n] = T\sus[n] + U[n] + E\suxc[n], \label{eq:hk_univ_f}
\end{equation}
where the non-interacting kinetic energy $T\sus[n]$ is the non-interacting limit of $F[n]$, and the electron-electron interaction contributes the familiar Hartree electrostatic interaction of the density with itself,
\begin{eqnarray}
  U[n] &=& \frac{1}{2}\int d\br \,  n(\br) u([n],\br) \label{eq:hartree_en} \\
  u([n],\br) &=& \int d\br' \frac{n(\br')}{|\br - \br'|}. \label{eq:hatree_pot}
\end{eqnarray}
The correction which makes Eq. \ref{eq:hk_univ_f} formally exact is the exchange-correlation (xc) energy $E\suxc[n]$.
Generalizing to spin-density functional theory and equating the Euler-Lagrange equations for the spin densities in the real and non-interacting systems implies that we can find the ground-state electron spin densities by self-consistently solving the one-electron Schr\"odinger equations
\begin{eqnarray}
  \left[-\frac{1}{2}\nabla^2 + v^{(\sigma)}\sus([\nup,\ndn],\br) \right] \varphi_{i\sigma}(\br) &=& \varepsilon_{i\sigma} \varphi_{i\sigma}(\br) \label{eq:ks_one_elec_se} \\
  v^{(\sigma)}\sus([\nup,\ndn],\br) &=& v_\sigma(\br) + u([n],\br) + \frac{\delta E\suxc}{\delta n_\sigma(\br)} \label{eq:ks_pot} \\
  n_\sigma(\br) &=& \sum_{i}|\varphi_{i\sigma}(\br)|^2 \theta(\mu - \varepsilon_{i\sigma}).
    \label{eq:one_elec_dens}
\end{eqnarray}
$\theta(x \geq 0)=1, \, \theta(x < 0)=0$ is the zero-temperature limit of the Fermi-Dirac distribution.
Of course,
\begin{equation}
  T\sus[\nup,\ndn] = \sum_{i\sigma}\int d\br \, \varphi_{i\sigma}^*(\br)
    \left(-\frac{1}{2}\nabla^2 \right) \varphi_{i\sigma}(\br)
    \theta(\mu - \varepsilon_{v\sigma}). \label{eq:ts}
\end{equation}
The Kohn-Sham orbitals and orbital energies in Eqs. \ref{eq:ks_one_elec_se}, \ref{eq:one_elec_dens}, and \ref{eq:ts} are \textit{implicit} functionals of the electron density, which bring many important details of the electronic shell structure into the largest components of the ground-state energy.
Without them, DFT would not have much practical application.
Even if we had the exact functional, the orbitals and orbital energies would not have a precise physical interpretation (except for the highest-occupied or partly-occupied orbital energy, which is \cite{levy1984} minus the first ionization energy).
But, the orbital energies often approximate single-particle energies well enough to be qualitatively useful.

The only quantity in this ground-state theory that must be approximated in practice is the xc energy $E\suxc[n]$.
For this, Kohn and Sham \cite{kohn1965} proposed the local density approximation (LDA):
\begin{equation}
  E\suxc^\text{LDA}[n] = \int d\br \, n(\br) \varepsilon\suxc^\text{unif}(n(\br)),
    \label{eq:exc_lda}
\end{equation}
where $\varepsilon\suxc^\text{unif}(n)$ is the xc energy per particle in an electron gas of uniform density $n$.
For this, they used an early estimate by Wigner \cite{wigner1934} but more accurate estimates were made by Singwi and collaborators \cite{singwi1970}.
In 1980, essentially exact values were found from a quantum Monte Carlo calculation \cite{ceperley1980} (and subsequently parameterized and confirmed in various ways \cite{vosko1980,perdew1981,perdew1992,sun2010,bhattarai2018}).
The local spin density approximation (LSDA) employs $\varepsilon\suxc^\text{unif}(\nup,\ndn)$.
Equation \ref{eq:exc_lda} was designed to be exact for a uniform density, or one that varies slowly over space.
It is also reminiscent of Slater's $X\alpha$ approximation \cite{slater1974}, which was originally intended to be a simplification of Hartree-Fock theory.
One of us (JPP) was told by Walter Kohn that he did not have high hopes for LDA, but expected it to be only a little more accurate than the Hartree approximation ($E\suxc = 0$) that was being widely used in condensed matter physics in 1965.
For about five years, there was little interest in the Kohn-Sham paper.
Then condensed matter physicists found that LDA was vastly better than the Hartree approximation for metal surface energies \cite{lang1970}, lattice constants, and phonon frequencies.
We now know that, while the xc energy is only a small part of the total energy of an atom, it is ``nature's glue'' \cite{kurth2000} that provides most of the binding of one atom to another.
Understanding why LSDA worked so much better than expected, as explained in Section 2, was a foundation for the development of better approximations.

There are several reasons why a local density approximation is better suited to the xc energy than to, say, the kinetic energy.
Besides those that will be discussed in Section 2, two are mentioned here.
(1) the exact exchange energy is, like the kinetic energy, an integral over Kohn-Sham orbitals, but it is a double (or Fock) integral that averages out more of the shell-structure oscillations of the orbitals than does the single integral of Eq. \ref{eq:ts} \cite{perdew2013}.
(2) In a single-center system, such as an atom, the exchange energy and the correlation energy are relatively local or short-ranged, because the density itself is \cite{cohen2001}.
But in a multi-center system, such as a molecule or solid, while the exchange and correlation energies are each more nonlocal or long-ranged, their most nonlocal or long-range contributions tend to cancel one another.

Electrons have a simple and well-known mutual Coulomb interaction, so many exact mathematical properties of the exact $E\suxc[n]$, or for its separate terms $E\sux[n]$ and $E\suc[n]$, have been derived to constrain the approximations.

The appearance of one-electron wavefunctions (orbitals) in Kohn-Sham theory can lead to the spurious conclusion that this theory, like Hartree-Fock theory \cite{szabo1982}, is only valid for ``uncorrelated'' or ``weakly-correlated'' electronic systems.
The exact Kohn-Sham theory should provide the exact ground-state density and energy (but not the exact excitation spectrum) for any electronic system, however strongly-correlated.
``Strong correlation'' is a challenge only for the approximations, and one that is starting to be met.
From our viewpoint, bonding electrons almost always display important correlation that is reasonably included for most systems, even in an LSDA calculation.
For example, consider the atomization energies of the six representative \textit{sp}-bonded molecules in the AE6 set \cite{lynch2003}.
The mean absolute error of the Hartree-Fock approximation is 6.3 eV; this error is reduced to 3.3 eV in LSDA, and to 0.1--0.6 eV or as little as 1\% error in good non-empirical generalizations of LSDA \cite{perdew2021}.

The exact exchange energies in the correlated wavefunction and density functional theories differ only via the typically small differences between Hartree-Fock and Kohn-Sham occupied orbitals.
The exchange and correlation energies are negative corrections to the Hartree approximation for the energy, arising for three reasons:
\begin{enumerate}
  \item An electron has a repulsive Coulomb interaction with other electrons but \textit{not} with itself.
  \item The Pauli exclusion principle, which operates even in the absence of interactions, keeps one electron from getting too close to others of the same spin.
  \item The Coulomb repulsion further tends to keep any two electrons apart.
\end{enumerate}
For a given density, these effects lower the expectation value of the third term on the right-hand side of Eq. \ref{eq:full_hamil} (and slightly increase the expectation value of the first term).
This dance of the electrons strengthens bonding because an electron acquires more neighbors with which to correlate by the Pauli or Coulomb mechanisms as the atoms get closer together.

\section{Exact constraints and a ladder of approximations}

Despite a surfeit of density functional approximations (DFAs), there is no unequivocal ``best'' DFA at present.
This section aims to give the reader the tools and references needed to analyze when and why DFAs work well.

To \textit{systematically} construct better, general-purpose DFAs, experience shows the best path forward to be incorporation of more behaviors known of the exact xc functional, or ``exact constraints'' for short.
One may also ensure that a DFA accurately describes simple, non-bonded systems, or ``appropriate norms.''
These types of DFAs are referred to as non-empirical or first-principles.
Fits to bonded systems rely on known but uncontrollable error cancellation between exchange and correlation DFAs, often resulting in numerically irregular DFAs of limited applicability.
DFAs that fit to bonded systems are termed empirical.

Thus it is crucial to emphasize the systematic construction of DFAs that better describe \textit{all} electronic matter.
There are only 17 known constraints that can be possibly satisfied by semi-local (SL) DFAs, which are of the form
\begin{equation}
  E\suxc^\text{SL}[\nup,\ndn] = \int \, e^\text{SL}\suxc(\nup,\ndn,...; \br) d\br,
\end{equation}
i.e., a single integral over real space of a quantity $e^\text{SL}\suxc$ defined by the spin densities and Kohn-Sham orbitals in an infinitesimal neighborhood of each position $\br$.
This necessarily requires that the model xc energy density $e\suxc^\text{SL}$ only depends on local quantities: the spin-densities $n_\sigma(\br)$, their spatial derivatives, and perhaps upon the local kinetic energy spin-densities (hence the ellipsis).
In contrast, non-local functionals, whose energy densities are typically two-point function(al)s, can satisfy four more exact constraints.
There are innumerably infinite ways to incorporate all 17 known exact constraints in a DFA, thus a systematic approach is needed to find the \textit{right} way.

Invariance of the energy under translation or rotation of the density in
empty space is trivially satisfied, and thus not counted among the exact  constraints.

\section{Expressions for the exact exchange and correlation energies}

In the original derivation of the Hohenberg-Kohn theorem, the energy as a functional of the density is defined only for densities that are ``pure-state $v$-representable,'' arising from non-degenerate ground-state wavefunctions for electronic systems in the presence of external potentials $v(\br)$.
But there are reasonable-looking densities that are not of this type, e.g., ground-ensemble densities constructed from degenerate ground-state wavefunctions \cite{levy1982}.

To define the energy functionals on the much-larger and better-characterized class of ``$N$-representable'' densities arising from any $N$-electron wavefunction, and to define the functionals more constructively, we turn to the constrained-search method of Percus \cite{percus1978}, Levy \cite{levy1979}, and Lieb \cite{lieb1983}.
Without the nucleus-nucleus Coulomb repulsion, the Hamiltonian of Eq. \ref{eq:full_hamil}, can be written as
\begin{equation}
  \hat{H}_\lambda = \hat{F}_\lambda + \sum_{i=1}^N v(\br_i),
\end{equation}
where $\hat{F}_\lambda = \hat{T} + \lambda \hat{V}_\text{ee}$ and
\begin{eqnarray}
  \hat{T} &=& -\frac{1}{2}\sum_{i=1}^N \nabla_i^2 \\
  \hat{V}_\text{ee} &=& \sum_{i=1}^N \sum_{j>i} |\br_i - \br_j|^{-1}.
\end{eqnarray}
The coupling constant $\lambda$ changes the strength of the Coulomb repulsion, $e^2 \to \lambda e^2$.
Then let $\Psi_n$ be the many-electron wavefunction that minimizes the expectation value of $\hat{F}_1$, at the physical value of the coupling constant, and $\Phi_n$ the many-electron wavefunction that minimizes the expectation value of $\hat{F}_0 = \hat{T}$, describing the fictional non-interacting (Kohn-Sham) system.
Both $\Psi_n$ and $\Phi_n$ are fixed to yield the same density $n(\br)$ through Eq. \ref{eq:spin_dens}, assuming that the interacting ground-state density is also ``non-interacting $v$-representable.''

Then the exchange energy is defined as \cite{levy1985}
\begin{equation}
  E\sux[n] \equiv \expec{\Phi_n}{\hat{V}_\text{ee}} - U[n], \label{eq:ex_def}
\end{equation}
with $U[n]$ given by Eq. \ref{eq:hartree_en}, and the correlation energy as \cite{levy1985}
\begin{eqnarray}
  E\suc[n] &\equiv & \expec{\Psi_n}{(\hat{T} + \hat{V}_\text{ee})}
    - \expec{\Phi_n}{(\hat{T} + \hat{V}_\text{ee})} \label{eq:ec_cons_search} \\
  &=& T\suc[n] + \expec{\Psi_n}{\hat{V}_\text{ee}} - U[n] - E\sux[n]. \nonumber
\end{eqnarray}
$ T\suc[n] = \expec{\Psi_n}{\hat{T}} - \expec{\Phi_n}{\hat{T}}$ is commonly called the ``kinetic energy due to correlation.''
These equations are easily generalized from density ($n$) to spin-density ($\nup,\,\ndn$) \cite{perdew1995}.

A complementary approach defines the xc and correlation energies through a coupling-constant integration \cite{harris1974} between the noninteracting ($\lambda=0$) and physical ($\lambda=1$) systems.
Under the constraint that $n(\br)$ is independent of $\lambda$ \cite{langreth1975,gunnarsson1976,langreth1977},
\begin{eqnarray}
  E\suxc[n] &\equiv & \int_0^1 d\lambda \expec{\Psi_n^{(\lambda)}}{\hat{V}_\text{ee}}
    - U[n] \label{eq:ac_exc} \\
  E\suc[n] &\equiv & \int_0^1 d\lambda \expec{\Psi_n^{(\lambda)}}{\hat{V}_\text{ee}}
    - \expec{\Psi_n^{(0)}}{\hat{V}_\text{ee}}. \label{eq:ac_ec}
\end{eqnarray}
The expression for $E\sux[n]$ does not change with $\lambda$, as the exchange energy involves expectation values that are taken in the non-interacting system only.
$\Psi_n^{(\lambda)}$ is the wavefunction that minimizes the expectation value of $\hat{F}_\lambda$ at each value of $\lambda$; thus $\Psi_n^{(0)}\equiv \Phi_n$ and $\Psi_n^{(1)}\equiv \Psi_n$.

Both sets of expressions (Eqs. \ref{eq:ex_def}-\ref{eq:ec_cons_search} and \ref{eq:ac_exc}-\ref{eq:ac_ec}) have been used to derive the exact constraints explicated here.
The constraints in the next section can be derived \cite{langreth1975,gunnarsson1976} from Eqs. \ref{eq:ac_exc}-\ref{eq:ac_ec}.

\subsection{The exchange and correlation holes}

Modern models of the xc energy typically approximate the xc energy per electron $\varepsilon\suxc = e\suxc/n$ directly.
Some of the first practical models of $\varepsilon\suxc$, such as those of Perdew and Wang \cite{perdew1985,perdew1986,perdew1991}, modeled the non-local (coupling-constant averaged) xc hole $n\suxc(\br,\br')$, computed
\begin{equation}
  \varepsilon\suxc(\br) = \frac{1}{2}\int d \br' \frac{n\suxc(\br,\br')}{|\br - \br'|}, \label{eq:xc_en_from_hole}
\end{equation}
and parametrized the results as functions of $n$ and $\nabla n$.
Equation \ref{eq:xc_en_from_hole}, which resembles Eq. \ref{eq:hatree_pot} for the Hartree potential, states that the xc energy arises from the interaction between an electron and the xc hole surrounding it.
Just as for the xc energy, the hole can be separated into distinct exchange and correlation contributions $n\suxc(\br,\br') = n\sux(\br,\br') + n\suc(\br,\br')$.
While Eq. \ref{eq:xc_en_from_hole} defines a particular xc energy density in the ``Hartree gauge,'' xc energy densities are neither measurable nor unique.

There are only six known constraints unique to the xc hole that cannot be expressed as constraints on the xc energy density.
The constraints discussed in this section can be satisfied by a local or semi-local hole model; a nonlocal hole constraint is discussed in Sec. \ref{sec:nonloc_cons}.
The transition from constructions based on hole constraints to those based on energy functional constraints occurred with the Perdew-Burke-Ernzerhof (PBE) \cite{perdew1996} generalized gradient approximation (GGA).
In fact, the same PBE GGA was found \cite{perdew1996a} from both constructions, and the hole construction guided PBE's choice between energy functional constraints that are incompatible at the GGA level (constraints \textbf{x4} and \textbf{xc4}).

\textbf{i.} The exchange hole integrates to one missing charge
\begin{equation}
  \int d\br' \, n\sux(\br,\br') = -1.
\end{equation}
The DFT exchange energy, like its Hartree-Fock analog, enforces the Pauli principle (or spin-$1/2$ spin-statistics theorem): no two electrons can possess identical quantum numbers.
Intuitively, this sum rule tells us that, if an electron of spin $\sigma$ is found at position $\br$, it is missing from the density of electrons of spin $\sigma$ around $\br$.

\textbf{ii.} The exchange hole is non-positive,
\begin{equation}
  n\sux(\br,\br') \leq 0. \label{eq:nx_nonpos}
\end{equation}
Properties \textbf{i} and \textbf{ii} can be derived from the definition \cite{harris1974} of the exact exchange hole
\begin{equation}
  n\sux(\br\sigma,\br'\sigma') = - \frac{\delta_{\sigma\sigma'}}{n(\br)}
  \left|\sum_i \phi^*_{i\sigma}(\br)\phi_{i\sigma}(\br')
    \theta(\mu - \varepsilon_{i\sigma})\right|^2.
  \label{eq:nx_def}
\end{equation}

\textbf{iii.}The on-top exchange hole density is
\begin{equation}
  n\sux(\br\sigma,\br\sigma)=-n_\sigma(\br)^2/n(\br).
\end{equation}
Note that $n\sux(\br,\br')= \sum_\sigma n\sux(\br\sigma,\br'\sigma)$.

\textbf{iv.} The correlation hole is charge neutral \cite{gunnarsson1977},
\begin{equation}
  \int d\br' \, n\suc(\br,\br') = 0. \label{eq:nc_sr}
\end{equation}
Thus correlation can alter the shape of the xc hole, but not the charge contained within it.

\textbf{v.} The exact, spherically-averaged xc hole has a cusp when $\br'=\br$ \cite{burke1998}.
Let $\widetilde{n}\suxc^{(\lambda)}(\br,\bm{u})=n\suxc^{(\lambda)}(\br,\br+\bm{u})$ be the xc hole at coupling-constant $\lambda$, such that $\widetilde{n}\suxc(\br,\bm{u}) = \int_0^1 d\lambda \, \widetilde{n}\suxc^{(\lambda)}(\br,\bm{u})$, and $\bm{u} = \br'-\br$.
Then its spherical average is
\begin{equation}
  \widetilde{n}^{(\lambda)}\suxc(\br,u) =
    \frac{1}{4\pi}\int d\Omega_{\bm{u}} \, \widetilde{n}\suxc^{(\lambda)}(\br,\bm{u}),
\end{equation}
and the cusp condition is \cite{burke1998}
\begin{equation}
  \lim_{u \to 0}\frac{d \widetilde{n}\suxc^{(\lambda)}(\br,u)}{d u} =
    \lambda \left[ n(\br) + \widetilde{n}\suxc^{(\lambda)}(\br,0) \right].
\end{equation}
The exchange hole, $n\suxc^{(0)}$, lacks a cusp as $u\to 0$.
Electrons in classical dynamics cannot come together at the same point, but quantum mechanical electrons can.
This gives rise to a non-zero probability density $n\suxc^{(\lambda)}(\br,\br) = n(\br) + \widetilde{n}\suxc^{(\lambda)}(\br,0)$ to find a second electron on top of an electron already at $\br$.

The LSDA hole satisfies all the hole constraints presented here, but the second-order gradient expansion hole (discused in the next section) requires long-range cutoffs to restore some of them.

\subsection{Constraints on the exchange energy alone}

\textbf{x1.} Because the exchange hole is non-positive (Eq. \ref{eq:nx_nonpos}), the exchange energy is negative definite
\begin{equation}
  E\sux[\nup,\ndn] < 0.
\end{equation}
To satisfy this condition for all possible densities, the exchange energy density $e\sux(\br)$ of a semi-local approximation must be non-positive.

\textbf{x2.} Consistent with the Pauli principle, the exchange energy is nonzero only for electrons of the same spin.
Given an exchange functional $E\sux[n]$ for a spin-unpolarized density $n(\br)$, the exchange energy for arbitrary spin-densities $\nup(\br)$ and $\ndn(\br)$ is \cite{oliver1979}
\begin{equation}
  E\sux[\nup,\ndn] = \frac{1}{2}\left( E\sux[2\nup] + E\sux[2\ndn] \right). \label{eq:x_spin_scl}
\end{equation}
To the best of our knowledge, no modern DFAs violate constraints \textbf{x1} and \textbf{x2}.

\textbf{x3.} Let $\gamma \geq 0$ be a control parameter and rescale the position vector $\br \to \br_\gamma = \gamma \br$ such that each coordinate is multiplied by $\gamma$.
The total number of electrons will be unchanged by this transformation only if the local density scales as
\begin{equation}
  n_\gamma(\br) = \gamma^3 n(\gamma \br).
\end{equation}
Using Eq. \ref{eq:spin_dens}, the Kohn-Sham orbitals scale as
\begin{equation}
  [\phi_{i\sigma}(\br)]_\gamma = \gamma^{3/2} \phi_{i\sigma}(\gamma \br),
\end{equation}
such that $[\phi_{i\sigma}(\br)]_\gamma$ is normalized if $\phi_{i\sigma}( \br)$ is.
Scaled quantities are denoted with a subscripted $\gamma$; to avoid confusion with other subscripts, we use the notation $[f_\alpha(\br)]_\gamma$ to indicate that $f_\alpha(\br)$ is scaled by $\gamma$, as needed.

\begin{figure}
  \includegraphics[width=0.7\columnwidth]{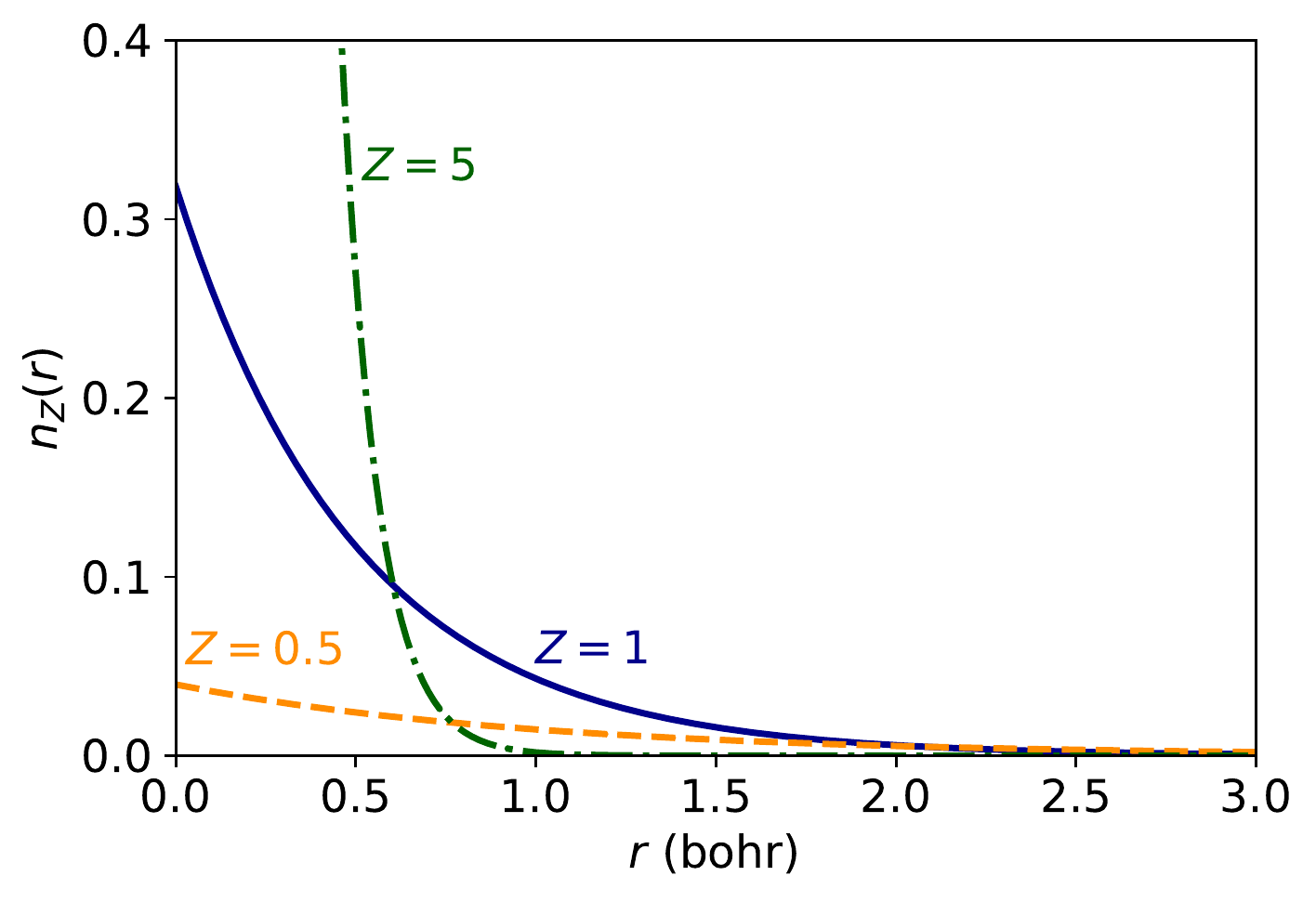}
  \caption{ Uniformly scaled hydrogen atom density $n_Z(r) = Z^3 e^{-2 Z r}/\pi$, with the origin at the nucleus.
  The uniform scaling parameter $Z$ mimics the nuclear charge of a non-interacting hydrogenic density.
  \label{fig:nh_us}}
\end{figure}

This ``uniform coordinate scaling'' transformation also preserves the total number of electrons.
For $\gamma > 1$, the density localizes towards the (in general, arbitrary) origin, tending to a spike as $\gamma \to \infty$.
For $\gamma < 1$, the density is lowered at the origin.
Figure \ref{fig:nh_us} demonstrates this using the exact hydrogen atom density.

Under uniform coordinate scaling, \cite{levy1985}
\begin{equation}
  E\sux[n_\gamma] = \gamma E\sux[n], \label{eq:x_unif_scl}
\end{equation}
i.e., the exchange energy is a homogeneous functional of the density with degree 1.
Using only this constraint, one can determine the functional form of the LDA for exchange,
\begin{equation}
  E\sux^\text{LDA}[n] = A\sux \int d\br \, n^{4/3}(\br),
\end{equation}
but not the coefficient $A\sux = -3(3\pi^2)^{1/3}/(4\pi)$ \cite{kohn1965}.
The generalization of LDA to arbitrary spin-densities using Eq. \ref{eq:x_spin_scl} for its exchange term, the local spin density approximation (LSDA), is the lowest rung on a ladder of approximations of generally increasing accuracy \cite{perdew2001}.

The next rung above the LSDA is the GGA, which also depends on the gradient of the density.
To construct a GGA that satisfies Eq. \ref{eq:x_unif_scl}, we deduce that
\begin{equation}
  s(\br) = \frac{|\nabla n(\br)|}{2\kf(\br) n(\br)}
\end{equation}
scales like $\gamma^0$ under uniform coordinate scaling, i.e., $s([n_\gamma];\br) =s([n];\gamma\br) $.
$s$ is a dimensionless gradient of the density on the length scale of the exchange energy, $2\pi/\kf$, where the local Fermi wave vector $\kf(\br) = [3\pi^2 n(\br)]^{1/3}$.
Then a GGA which satisfies Eq. \ref{eq:x_unif_scl} is
\begin{equation}
  E\sux^\text{GGA}[n] = A\sux \int d\br \, F\sux(s) n^{4/3}(\br).
\end{equation}
The exchange enhancement factor $F\sux$ modulates the LDA exchange energy as a function of the density inhomogeneity.
$s$ is often used in non-empirical DFAs, whereas $x=2(3\pi^2)^{1/3}s$ is often used in empirical DFAs.

The meta-GGA is the next rung above the GGA, and can further depend on the Laplacian of the density or the local kinetic energy density.
A Laplacian-level meta-GGA (or MGGA-L) that satisfies Eq. \ref{eq:x_unif_scl} is then
\begin{equation}
  E\sux^\text{MGGA-L}[n] = A\sux \int d\br \, F^\text{MGGA-L}\sux(s,q) n^{4/3}(\br),
\end{equation}
with $q$ a dimensionless Laplacian on the length scale of exchange,
\begin{equation}
  q(\br) = \frac{\nabla^2 n(\br)}{4 \kf^2(\br) n(\br)}.
\end{equation}
Laplacian-level meta-GGAs are better suited to metallic condensed matter \cite{kaplan2022}.

The more common orbital-dependent meta-GGAs are constructed from the non-negative kinetic energy density
\begin{equation}
  \tau_\sigma(\br) =  \frac{1}{2}\sum_i | \nabla\phi_{i\sigma}(\br)|^2
    \Theta(\mu - \varepsilon_{i\sigma}).
\end{equation}
The conventional one-body kinetic energy density
\begin{equation}
  \widetilde{\tau}_\sigma(\br) = -\frac{1}{2}\sum_i \phi_{i\sigma}^*(\br)
    \nabla^2 \phi_{i\sigma}(\br) \Theta(\mu - \varepsilon_{i\sigma})
\end{equation}
is, like $q$, unbounded.
As $\tau_\sigma - \text{Re}\, \widetilde{\tau}_\sigma = \nabla^2 n_\sigma/4$, $\tau_\sigma$ and $\widetilde{\tau}_\sigma$ yield the same integrated kinetic energy by the divergence theorem.

There are many dimensionless kinetic energy variables that preserve the uniform scaling constraint \cite{furness2019}.
Just as $|\nabla n|$ and $n^{4/3}$ have the same uniform coordinate scaling behavior, one can find semi-local variables with the same uniform scaling behavior as $\tau$ from its limiting behaviors.
$\tau$ is bounded from below by the von Weizs\"acker kinetic energy density \cite{hoffmann1977}
\begin{equation}
  \tau(\br) \geq \tau_\text{W}(\br) = \frac{|\nabla n(\br)|^2}{8n(\br)}.
\end{equation}
For a uniform density, $\tau$ tends to the Thomas-Fermi kinetic energy density
\begin{equation}
  \tau_\text{unif}(\br) = \frac{3}{10}\kf^2(\br) n(\br).
\end{equation}
The non-negative ``iso-orbital indicator''
\begin{equation}
  \alpha(\br) = \frac{\tau(\br) - \tau_\text{W}(\br)}{\tau_\text{unif}(\br)}
\end{equation}
tends to zero for one- and two-electron densities, and tends to one for a uniform density.
For weakly inhomogeneous densities, $|\alpha -1 | \ll 1$ \cite{brack1976}.
$\alpha$ was first proposed as an ingredient of the electron localization function (ELF) \cite{becke1990},
\begin{equation}
  \text{ELF}(\br,\sigma) = \frac{1}{1 + \alpha^2_\sigma(\br)}.
\end{equation}
$\alpha$ can identify different chemical bonds: $\alpha=0$ corresponds to a single-orbital bond; $0 < \alpha \lesssim 1$ to a covalent bond; $\alpha\approx 1$ to a metallic bond; and $\alpha \gg 1$ to a weak bond \cite{becke1990}.
Then a meta-GGA satisfying the uniform scaling constraint of Eq. \ref{eq:x_unif_scl} could be
\begin{equation}
  E\sux^\text{MGGA}[n] = A\sux \int d\br \, F\sux^\text{MGGA}(s,\alpha) n^{4/3}(\br).
\end{equation}

\textbf{x4.} For densities that vary sufficiently slowly in space ($s \ll 1$ and $|q| \ll 1$), the exchange energy tends to a ``density-gradient expansion'' \cite{svendsen1995,svendsen1996}
\begin{equation}
  E\sux[n] = A\sux \int d\br \left\{1 + \frac{10}{81}s^2 + \frac{146}{2025}\left( q^2 - \frac{5}{2}q s^2\right) + \mathcal{O}(|\nabla n|^6) \right\} n^{4/3}(\br). \label{eq:gex4}
\end{equation}
All higher-order terms are presently unknown.
The coefficients of $s^2$ and $q^2$ are known exactly.
The coefficients of $qs^2$ (-73/405) and $s^4$ (zero) are known with some uncertainty; we present the best estimates from Ref. \cite{svendsen1996}.

The precursor to the GGA, the gradient expansion approximation (GEA) \cite{kohn1965}, truncates Eq. \ref{eq:gex4} at finite order.
When $s \ll 1$, the GEA can describe the weak inhomogeneities typical of valence electron densities in simple metals, like the alkalis.
The GEA is a poor approximation for real atoms, molecules, and solids, which have $0 \leq s \lesssim 3$ in energetically relevant regions \cite{zupan1997}.
Equation \ref{eq:gex4} is derived only for systems without classical turning surfaces \cite{kaplan2021}, such as metals or small-gap semiconductors.

By construction, the LDA can only recover the leading-order term, $A\sux n^{4/3}$.
A GGA can recover the second-order term $10s^2/81$ in addition to the LDA term.
However, the best-known exchange GGAs, due to Perdew, Burke and Ernzerhof (PBE) \cite{perdew1996} and to Becke (B88) \cite{becke1988}, do not.
PBE (B88) finds a second-order gradient coefficient that is about 1.78 (2.22) times too large.
The value of PBE's second-order gradient coefficient was fixed by constraint \textbf{xc4}, which is incompatible with this constraint at the GGA level.
B88 was constructed to accurately reproduce the Hartree-Fock exchange energies of closed-shell atoms.
Their larger second-order coefficients improve their predictions of atomic and molecular properties \cite{elliott2009,cancio2018} at the expense of those for solids \cite{perdew2008}.
The Perdew-Wang 1991 (PW91) \cite{perdew1991} and PBEsol \cite{perdew2008} GGAs recover the second-order gradient expansion for exchange.

A Laplacian-level meta-GGA can recover the fourth-order expansion straightforwardly.
A $\tau$-level meta-GGA can also recover the fourth-order expansion using the gradient expansion of $\tau$ \cite{brack1976} and an integration-by-parts procedure \cite{aschebrock2019,furness2022}.
Modern non-empirical meta-GGAs, e.g., the Tao-Perdew-Staroverov-Scuseria (TPSS) \cite{tao2003}, revTPSS \cite{perdew2009}, strongly-constrained and appropriately normed (SCAN) \cite{sun2015}, and TASK \cite{aschebrock2019} meta-GGAs, recover this constraint to fourth-order.

Other non-empirical meta-GGAs, like the r$^2$SCAN \cite{furness2020}, recover only the second-order gradient expansion for exchange.
r$^2$SCAN was designed to improve primarily upon the numerical efficiency of SCAN, but makes slightly larger errors in solid-state properties and smaller errors in molecular properties than SCAN.
This is partly due its recovery of a lower-order gradient expansion for exchange \cite{furness2022}.

Most empirical DFAs -- for example, the M06-L \cite{zhao2006} and the machine-learned xc-diff \cite{dick2021} meta-GGAs -- recover only the leading-order LDA term.

\textbf{x5.} Rather than scale each coordinate by the same parameter as in constraint \textbf{x3}, consider scaling only one coordinate, say $x$,
\begin{eqnarray}
  [\phi_{i\sigma}(\br)]^x_\gamma &=& \gamma^{1/2} \phi_{i\sigma}(\gamma x, y, z) \\
  n^x_\gamma(\br) &=& \gamma n(\gamma x, y, z),
\end{eqnarray}
where again $\gamma \geq 0$.
This one-dimensional, non-uniform coordinate scaling preserves both the normalization of the Kohn-Sham orbitals and the total number of electrons.
Under such a one-dimensional scaling, the exact exchange energy per electron remains finite \cite{levy1991,gorling1992}
\begin{eqnarray}
  \lim_{\gamma \to \infty} \frac{1}{N}E\sux[n_\gamma^x] & > & - \infty \label{eq:ex_nus_hd} \\
  \lim_{\gamma \to 0} \frac{1}{N}E\sux[n_\gamma^x] & > & - \infty.
\end{eqnarray}
The $\gamma \to \infty$ limit describes the collapse of a system from three to two dimensions.

The high-density, $\gamma \to \infty$, constraint is seldom enforced: the LDA, which diverges as $\gamma^{1/3}$, and most higher-level approximations violate this constraint.
To enforce the constraint of Eq. \ref{eq:ex_nus_hd} \cite{chiodo2012,perdew2014}
\begin{equation}
  \lim_{s\to\infty}F\sux(s,...) \propto s^{-1/2}. \label{eq:fx_nonunif}
\end{equation}
This constraint is enforced by the MVS, SCAN, r$^2$SCAN, and TASK meta-GGAs.
Most DFAs satisfy the low-density $\gamma \to 0$ limit, where $E\sux \to 0^-$.

\textbf{x6.} Lieb and Oxford \cite{lieb1981} demonstrated that the indirect part
\begin{equation}
  I[n] = \expec{\Psi_n}{\hat{V}_\text{ee}} - U[n] = E\suxc[n] - T\suc[n], \label{eq:indirect}
\end{equation}
of the electron-electron repulsion for any non-relativstic, Coulombic system is bounded from below as
\begin{equation}
  I[n] \geq 2.273 E\sux^\text{LDA}[n], \qquad \text{any}~N.
\end{equation}
Using constraint \textbf{c1} for the non-positivity of the correlation energy, Perdew \cite{perdew1991} extended this result to the exchange and xc energies for a spin-unpolarized density $n$
\begin{equation}
  E\sux[n/2,n/2] \geq E\suxc[n/2,n/2] \geq 1.804 E\sux^\text{LDA}[n].
\end{equation}
Then the exchange enhancement factor should be bounded as $F\sux \leq 1.804$.
This bound is satisfied by many non-empirical exchange DFAs (e.g., LDA, PBE, TPSS, SCAN, r$^2$SCAN).
However many empirical exchange DFAs, like B88 [$F\sux(s\to \infty) \propto s/\ln s$] and M06-L [$F\sux(s\to\infty,\alpha) = 8.026$] badly violate this bound.

\textbf{x7.}  For any one-electron density $n_1$, an optimal lower bound on the indirect part of the Coulomb repulsion is known \cite{gadre1980,lieb1981}
\begin{equation}
  I[n_1] \geq 1.479 E^\text{LDA}\sux[n_1]. \label{eq:1e_lo}
\end{equation}
Using Eqs. \ref{eq:indirect} and \ref{eq:1e_lo}, Perdew \textit{et al.} \cite{perdew2014} derived a tight upper-bound on the exchange enhancement factor of a two-electron, spin-unpolarized density $n_2$
\begin{equation}
  F\sux(s,\alpha=0) \leq 1.174, \label{eq:2e_lo}
\end{equation}
and conjectured that this bound holds for $\alpha \geq 0$ in all systems.
Recent work has found a rigorous bound $F\sux(s,\alpha) \leq 1.3423$ when the wavefunction can be written as a single Slater determinant \cite{lewin2022}, bolstering this conjecture.
LDA exchange trivially satisfies this bound ($F^\text{LDA}\sux=1$).
Some meta-GGAs, like MVS, SCAN, r$^2$SCAN, and TASK enforce the rigorous bound.

\subsection{Constraints on the correlation energy alone}

\textbf{c1.} The correlation energy is non-positive \cite{levy1979,levy1985}
\begin{equation}
  E\suc[\nup,\ndn] \leq 0.
\end{equation}
This constraint is nearly-always enforced by first-principles DFAs, but is violated locally \cite{tran2016} by the Lee-Yang-Parr (LYP) \cite{lee1988,miehlich1989} correlation GGA.

\textbf{c2.} The correlation energy is identically zero for a one-electron density \cite{perdew1981}.
This constraint cannot be satisfied by an LSDA or GGA without making the correlation energy zero for any fully spin-polarized density, as LYP does.

For a one-electron density, the iso-orbital indicator $\alpha$ vanishes.
PKZB, TPSS, revTPSS, M06-L, SCAN and r$^2$SCAN use this fact to satisfy constraint \textbf{c2}.

\textbf{c3.} The correlation energy satisfies nontrivial uniform coordinate scaling constraints.
We consider two distinct limits: the high density limit $\gamma \to \infty$, and the low-density limit $\gamma \to 0$ (constraint \textbf{c4}).

Under uniform coordinate scaling to the high-density limit \cite{levy1989,levy1991}, in the absence of degeneracies,
\begin{equation}
  \lim_{\gamma \to \infty} E\suc[n_\gamma] > -\infty. \label{eq:ec_uscl_hd}
\end{equation}
Thus the correlation energy remains finite even for highly localized densities.
As the noninteracting kinetic energy scales with $\gamma^2$ \cite{levy1985}, a tendency to over-localize is prevented by a large positive increase in the kinetic energy.
Moreover, Levy and Perdew showed that \cite{levy1985}
\begin{equation}
  E\suc[n_\gamma] > \gamma E\suc[n], \qquad \gamma > 1.
\end{equation}

The LSDA, which logarithmically diverges as $\gamma \to \infty$ \cite{gell-mann1957}, does not satisfy Eq. \ref{eq:ec_uscl_hd}.
PBE, TPSS, SCAN, and r$^2$SCAN make the correlation energy scale to a negative constant as $\gamma \to \infty$ in an inhomogeneous system.

\textbf{c4.} Under uniform coordinate scaling to the low-density limit \cite{levy1993}
\begin{equation}
  \lim_{\gamma \to 0} E\suc[n_\gamma] = \gamma D\suc[n],
\end{equation}
where the density functional $D\suc[n]$ is unknown generally.
In the low-density limit, the Hartree \cite{levy1985}, exchange, and correlation energies all have the same scaling behavior, and tend to zero less rapidly than the non-interacting kinetic energy.
Moreover \cite{levy1985}
\begin{equation}
  E\suc[n_\gamma] < \gamma E\suc[n], \qquad \gamma < 1.
\end{equation}

To satisfy this constraint, an approximate correlation energy density must scale as $n_\gamma^{4/3}$ as $\gamma \to 0$.
The LSDA inherits the correct scaling behavior from its appropriate norm \cite{wigner1934}; PBE, SCAN, and r$^2$SCAN satisfy this constraint by construction.

\textbf{c5.} The correlation energy has a gradient expansion analogous to the exchange energy, commonly approximated as
\begin{equation}
  E\suc[\nup,\ndn] = \int d\br \left\{  \varepsilon\suc^\text{LSDA}(\rs,\zeta)
    + \beta(\rs) \phi^3(\zeta) t^2(\br)  + \mathcal{O}(|\nabla n|^4) \right\}\,n
    \label{eq:ec_ge}
\end{equation}
$t$ is a dimensionless gradient of the density on the length scale of the correlation energy: the Thomas-Fermi screening wavevector $k_\mathrm{s} = \sqrt{4\kf/\pi}$,
\begin{equation}
  t(\br) = \frac{|\nabla n|}{2 \phi(\zeta)k_\mathrm{s}(\br) n(\br) }.
\end{equation}
The local Wigner-Seitz radius $\rs(\br) = [4\pi n(\br)/3]^{-1/3}$ defines a sphere that contains one electron on average.
The relative spin-polarization $\zeta = (\nup - \ndn)/(\nup + \ndn)$ is the ratio of the magnetization density to the total density, bounded as $0 \leq |\zeta| \leq 1$.
The gradient coefficient $\beta(\rs)$ is weakly density dependent, with many possible parameterizations \cite{ma1968,wang1991,perdew1996,perdew2009,cancio2018}.
Its high-density limit is finite, $\beta(0)\approx 0.066725$ \cite{ma1968}.
Equation \ref{eq:ec_ge} neglects $\nabla \zeta \cdot \nabla n$ and $|\nabla \zeta|^2$ terms \cite{rasolt1981,wang1991}, but is exact for spin-unpolarized ($\zeta=0$) and fully spin-polarized densities ($|\zeta|=1$), and a reasonable approximation for $0 \leq |\zeta| \lesssim 0.6$ and $0.85 \lesssim |\zeta| \leq 1$ \cite{furness2022}.

Truncating Eq. \ref{eq:ec_ge} at second order yields the correlation GEA.
The correlation GEA often over-corrects the LSDA for slowly-varying systems, such as simple metal surfaces \cite{langreth1980}, and finds positive correlation energies for atoms \cite{ma1968,cancio2018}, violating constraint \textbf{c1}.
Further, the underlying GEA correlation hole violates the sum rule of Eq. \ref{eq:nc_sr} \cite{langreth1980}.

By definition, the LSDA can only recover the leading-order term in Eq. \ref{eq:ec_ge}.
PW91 and PBE recover the high-density limit of this expression, whereas PBEsol's gradient coefficient is $0.689\beta(0)$.
meta-GGAs like revTPSS, SCAN, and r$^2$SCAN recover an $\rs$-dependent gradient coefficient.
It is uncommon for empirical DFAs to recover even the leading-order LSDA term (e.g., LYP), however M06-L does by constraint.
Although the LSDA correlation energy in finite systems is too negative, LSDA exchange and correlation make complementary errors such that their sum is much more accurate \cite{burke1998}.

\textbf{c6.} The correlation energy per electron satisfies similar non-uniform coordinate scaling constraints as those for the exchange energy \cite{gorling1992,pollack2000}
\begin{eqnarray}
  \lim_{\gamma \to \infty} \frac{1}{N}E\suc[n_\gamma^x] & > & - \infty \\
  \lim_{\gamma \to 0} \frac{1}{N}E\suc[n_\gamma^x] & = & 0.
\end{eqnarray}
These constraints can be satisfied by setting the large-$t$ limit of the correlation energy to zero.
As this is already done to enforce the high-density limit of uniform coordinate scaling, GGAs like PBE and PBEsol, meta-GGAs like TPSS, revTPSS, SCAN, and r$^2$SCAN all satisfy this constraint.
The LSDA diverges logarithmically as $\gamma \to \infty$, but tends to zero as $\gamma \to 0$.

\subsection{Constraints on the exchange-correlation energy}

\textbf{xc1.} Consider a system with distinct, well-separated subsystems A and B.
At sufficient separation, the electron density in subsystem A, $n_\text{A}$, and in subsystem B, $n_\text{B}$, should not overlap, and the total xc energy will be the sum over two independent subsystems
\begin{equation}
  E\suxc[n_\text{A} + n_\text{B}] = E\suxc[n_\text{A}] + E\suxc[n_\text{B}]. \label{eq:size_cons}
\end{equation}
This ``size-consistency'' \cite{pople1976} principle is also true for the total energy.
This property of the energy makes intuitive physical and chemical sense: electrons in isolated systems should not influence each other.
This constraint is satisfied by nearly all DFAs.
A DFA that depends explicitly upon the number of electrons, such as the Fermi-Amaldi approximation \cite{ayers2005}, will not be size-consistent.
The correlated-wavefunction configuration interaction (CI) method is not size consistent \cite{pople1976} except in the full CI limit.

Most semi-local DFAs need to break spin-symmetries to dissociate molecules into neutral, independent fragments \cite{perdew2021} consistent with Eq. \ref{eq:size_cons}.
Consider the dissociation of H$_2$ \cite{gunnarsson1976}: a spin-restricted LSDA calculation for H$_2$ at large separations found that the energy of the spin-unpolarized molecule was much greater than the energy of two isolated H atoms.
A spin-unrestricted LSDA calculation found a quantitatively correct dissociation, but the molecule had net spin up on one atom, and net spin down on the other -- a broken spin-symmetry configuration with zero net $z$-component of spin.
A fully non-local xc hole is needed to describe the symmetry-unbroken dissociation of a molecule or solid.

\textbf{xc3.} At sufficiently low densities, the xc energy is almost independent of the degree of spin-polarization $\zeta$ \cite{seidl2000,tao2003}
\begin{equation}
  \lim_{\gamma\to 0} E\suxc[n_{\uparrow\gamma},n_{\downarrow\gamma}]
    \approx E\suxc[n_\gamma].
\end{equation}
A seminal calculation \cite{ceperley1980} of the low-density phases of the UEG found that, at $\rs = 50$ bohr, the difference in ground-state energies between the paramagnetic ($\zeta=0$) and ferromagnetic ($|\zeta|=1$) fluid phases was about 1.63 meV in magnitude, and only 0.52 meV at $\rs=100$ bohr.
This weak dependence on $\zeta$ in the low-density fluid phase was reconfirmed recently \cite{holzmann2020}.
(The fluid phases are spherically symmetric, whereas the Wigner crystal phase breaks spherical symmetry by forming a close-packed lattice of electrons).

This constraint is enforced by TPSS, SCAN, and r$^2$SCAN.
LSDA inherits this near $\zeta$-independence at low densities from its appropriate norm, the UEG.
Empirical DFAs like LYP do not, because $E\suc^\text{LYP}=0$ whenever $|\zeta|=1$; however M06-L may approximate this $\zeta$-independence, as it reduces to the LSDA for homogeneous densities.

\textbf{xc4.} The linear response of a UEG to a low-energy, static (constant in time) perturbation is well-described by the LSDA linear response function for wave-vectors $k \lesssim 2\kf$ \cite{moroni1995}.
To mimic the LSDA linear response in a gradient-corrected DFA requires cancellation between the linear-response inhomogeneity contributions from exchange and correlation \cite{tao2008}.
The LSDA and PBE satisfy this constraint by construction, while approximations that recover the second-order gradient expansion for correlation and the fourth-order gradient expansion for exchange, such as TPSS and SCAN, come close to satisfying it at high and valence-electron densities.

\textbf{xc5.} For \textit{any} two-electron density $n_2$, Lieb and Oxford derived a lower bound on the indirect energy $I[n_2]$, which we cast as \cite{lieb1981,sun2016}
\begin{equation}
  E\suxc[n_2] \geq 1.671 E\sux^\text{LDA}[n_2].
\end{equation}
This bound can only be enforced in meta-GGAs such as SCAN and r$^2$SCAN, which can differentiate uniquely between one- and two-electron densities ($\alpha=0$) and all others ($\alpha>0$).

\subsection{Nonlocal constraints \label{sec:nonloc_cons}}

\textbf{n1.} The xc hole, at any value of the coupling-constant $\lambda$, represents a deficit in the density of electrons, \cite{burke1997}
\begin{equation}
  n\suxc^{(\lambda)}(\br,\br') \geq -n_\sigma(\br'). \label{eq:xc_hole_bd}
\end{equation}
Further, because $n\sux(\br,\br') \equiv n\suxc^{(0)}(\br,\br')$ \cite{burke1997},
\begin{equation}
  n\sux(\br,\br') \geq -n_\sigma(\br').
\end{equation}
The xc and exchange holes cannot remove more electrons at $\br'$ than are present at $\br'$.
A semi-local approximation to the xc hole can only rigorously fulfill this constraint for $\br = \br'$, or for systems where the hole model is exact (like the UEG).

\textbf{n2.} For any one- ($n_1$) or two-electron ($n_2$) density, Eq. (\ref{eq:nx_def}) for the exact exchange hole can be used to show that
\begin{equation}
  E\sux[n_N] = -\frac{1}{N} U[n_N], \qquad N=1,2.
\end{equation}
The Hartree electrostatic energy counts the interaction of every electron with itself, a one-electron self-interaction error.

As the Hartree potential is a nonlocal functional of the density, semi-local (SL) density functionals make one-electron self-interaction errors, which can be corrected as \cite{perdew1981}
\begin{equation}
  E\suxc^\text{SL-SIC}[\nup,\ndn] = E\suxc^\text{SL}[\nup,\ndn] - \sum_{i,\sigma} \left\{ U[n_{i\sigma}] + E\suxc^\text{SL}[n_{i\sigma},0] \right\}. \label{eq:pz_sic}
\end{equation}
Here, $n_{i\sigma} = f_{i\sigma}|\phi_{i\sigma}|^2$ is an orbital density with occupancy $0 \leq f_{i\sigma} \leq 1$.
The self-interaction correction (SIC) vanishes for the exact xc functional.

The Perdew-Zunger (PZ) SIC \cite{perdew1981} of Eq. \ref{eq:pz_sic} loses \cite{santra2019} the correct uniform density limit and is dependent upon the choice of orbital representation.
Thus PZ-SIC is size-inconsistent unless localized orbitals are employed.
Two sets of orbitals which are linked by a unitary transformation (one that leaves the density and each term in the total energy invariant) can give different SICs.
There are various systematic methods to construct a unitary transformation of the canonical orbitals, for example, the Fermi-L\"owdin orbital (FLO) SIC \cite{pederson2014}.

A carefully-constructed SIC can dramatically improve the accuracy of a DFA.
For a set of six molecular atomization energies, the LSDA makes a mean absolute error (MAE) of 74.26 kcal/mol; applying the SIC of Eq. \ref{eq:pz_sic} to the LSDA reduces the MAE to 57.97 kcal/mol; constraining the SIC to preserve the UEG limit further reduces the MAE to 9.95 kcal/mol, below that of PBE (13.43 kcal/mol) \cite{zope2019}.

Meta-GGAs like TPSS, SCAN, and r$^2$SCAN are self-correlation free and need a SIC only for their exchange energies.
Hartree-Fock theory does not make one-electron self-interaction errors.

The next rung above meta-GGAs are hybrid DFAs, which mix SL and single-determinant exchange (exact exchange, EXX, for a system with no near degeneracies),
\begin{eqnarray}
  e\sux^\text{EXX}(\br) &=& -\frac{1}{2} \sum_\sigma \int
  \frac{\left|\sum_i \phi^*_{i\sigma}(\br)\phi_{i\sigma}(\br')
  \theta(\mu - \varepsilon_{i\sigma})\right|^2 }{|\br - \br'|} d\br' \\
  e\suxc^\text{hybrid}(\br) &=& e\suxc^\text{SL}(\br)
    + a \left[e\sux^\text{EXX}(\br) - e\sux^\text{SL}(\br) \right] \label{eq:exc_glob_hyb}
\end{eqnarray}
Any fraction of EXX is compatible with the exact constraints discussed here, however, there is no universal constant mixing parameter $a$.
Hybrids make a self-interaction error through their SL components.
Global hybrids use a constant mixture of EXX and have energy densities of the form of Eq. \ref{eq:exc_glob_hyb}.
Perhaps the most famous global hybrids are PBE0 \cite{adamo1999} ($a=0.25$) and B3LYP \cite{stephens1994} (which uses three mixing parameters).
Range-separated hybrids, such as Heyd-Scuseria-Ernzerhof 2006 (HSE06) \cite{heyd2003} and $\omega$B97X-V \cite{mardirossian2014}, build upon the global hybrid form by screening the long-range part of EXX.
Local hybrids, like the range-separated DeepMind 21 (DM21) \cite{kirkpatrick2021}, use a position-dependent mixing of semi-local and exact exchange.

\textbf{n3.} When viewed as a function of \textit{fractional} electron number $\widetilde{N}$, the exact xc energy is a piecewise linear function of $\widetilde{N}$ \cite{perdew1982}.
Let $\widetilde{N} = N + M$, where $N \geq 1$ is a positive integer, and $-1 < M < 1$.
For $-1 < M < 0$, $\partial E\suxc/\partial \widetilde{N}$ in an isolated atom is minus the ionization potential, and for $0 < M < 1$, $\partial E\suxc/\partial \widetilde{N}$ is minus the electron affinity.
When $M=0$, $\partial E\suxc/\partial \widetilde{N}$ may be undefined; the difference
\begin{equation}
  \Delta\suxc = \lim_{\eta \to 0^+} \left[ \left. \frac{\partial E\suxc}{\partial \widetilde{N}}\right|_{N+\eta}
  - \left. \frac{\partial E\suxc}{\partial \widetilde{N}}\right|_{N-\eta} \right]
\end{equation}
is called the xc derivative discontinuity.

A fractional electron number can arise as the ensemble average of the density in an open system over a finite amount of time \cite{perdew1982}.
An erroneous fractional electron number can manifest as a charge-transfer error from a DFA.
All SL DFAs average over the non-analytic behavior of $E\suxc(\widetilde{N})$, making it a smooth function of $\widetilde{N}$ \cite{perdew2021a}.
The smoothed $E\suxc^\text{SL}(\widetilde{N})$ can minimize with a non-zero charge transfer towards one nuclear center.
The piecewise linearity can be restored with a SIC \cite{perdew2021a}, or by a localized orbital scaling correction (LOSC) \cite{li2017}.

The xc derivative discontinuity contributes to the fundamental bandgap $E_\text{gap}$ of a solid \cite{yang2016,perdew2017}
\begin{equation}
  E_\text{gap} = E_\text{gap}^\text{KS} + \Delta\suxc.
\end{equation}
The LSDA and GGAs have $\Delta\suxc = 0$ identically, but approximate the Kohn-Sham bandgap $E_\text{gap}^\text{KS}$ well, thus underestimating the fundamental gap.
Orbital-dependent DFAs, like meta-GGAs or hybrids, also have $\Delta\suxc  =0 $, but are implemented in a generalized Kohn-Sham (GKS) approach in which their exchange-correlation potentials are differential or integral operators.
Only in a GKS theory can we hope for a realistic orbital-energy gap between the valence and conduction bands \cite{perdew2017}.
Notably, the TASK meta-GGA \cite{aschebrock2019} has been constructed with the objective of predicting more accurate bandgaps.
Thus bandgap underestimation by semi-local DFAs reflects a limitation of the level of approximation and not of GKS DFT.

\textbf{n4.} The xc energy, viewed as a function of the fractional $z$-component of the electron spin, should make the total energy invariant \cite{cohen2008}.
Approximations that respect this constraint do not break spin symmetry.

A ground-ensemble density composed of multiple pure states can yield an effective fractional total spin.
A correction for semi-local DFAs similar to LOSC has been developed \cite{su2018} to restore this constraint.
The machine-learned DM21 \cite{kirkpatrick2021} local hybrid was taught to (nearly exactly) obey constraints \textbf{n3} and \textbf{n4} \cite{kirkpatrick2021}, and is exceptionally accurate for describing molecular chemistry \cite{gould2022}.
The loss of symmetry breaking due to satisfaction of constraint \textbf{n4} is not an unmixed blessing.
Symmetry breaking in DFT can reveal strong correlations and long-lived fluctuations \cite{perdew2021} (including antiferromagnetism) that become increasingly observable as a system becomes more extended over space.

\section{Ascending the ladder in real systems}

The Jacob's ladder hierarchy \cite{perdew2001} of Perdew and Schmidt is a common categorization scheme for DFAs that aims to understand how DFAs improve as they grow more mathematically sophisticated.
To ascend the ladder, one adds more variables to a DFA, thereby permitting a greater degree of nonlocality.
A higher-rung DFA \textit{can} be more accurate for a wider variety of densities than a lower-rung DFA.

The ladder rests on the Hartree world, but even its first rung, the LSDA, is typically more accurate than approximations that omit Coulomb correlation.
The Kohn-Sham exchange-only approximation (EXOA), which takes $E\suxc\approx E\sux^\text{EXX}$, is similar but inequivalent to the Hartree-Fock approximation (HFA), insomuch as the Kohn-Sham and Hartree-Fock wavefunctions differ \cite{langreth1983}.
As the differences are generally small in practice and the EXOA is less commonly used than the HFA, we will use Hartree-Fock energies as a stand-in for the EXOA when needed.
While EXX is an ingredient of more sophisticated DFAs, the EXOA neglects the correlation energy in its entirety.

The lowest rung on the ladder, the LSDA, which describes the exact exchange and correlation energies of an interacting uniform electron gas, is a more realistic approximation than the EXOA.
The average errors in atomization energies (the energy needed to separate a molecule into its atomic constituents) of first- to third-row molecules are reduced by about half in the LSDA from the EXOA \cite{kurth1999} or the HFA \cite{neumann1996}.
The average LSDA atomization energy error is still on the order of tens of kcal/mol, far from the threshold of chemical accuracy (conventionally less than 1 kcal/mol in absolute error).
The LSDA also tends to overbind first- to third-row molecules, whereas the EXOA or HFA tend to underbind them.
Molecular bond lengths are also generally improved by the LSDA, although differences from the HFA are less pronounced \cite{neumann1996}.

\begin{figure}
  \centering
  \includegraphics[width=0.8\columnwidth]{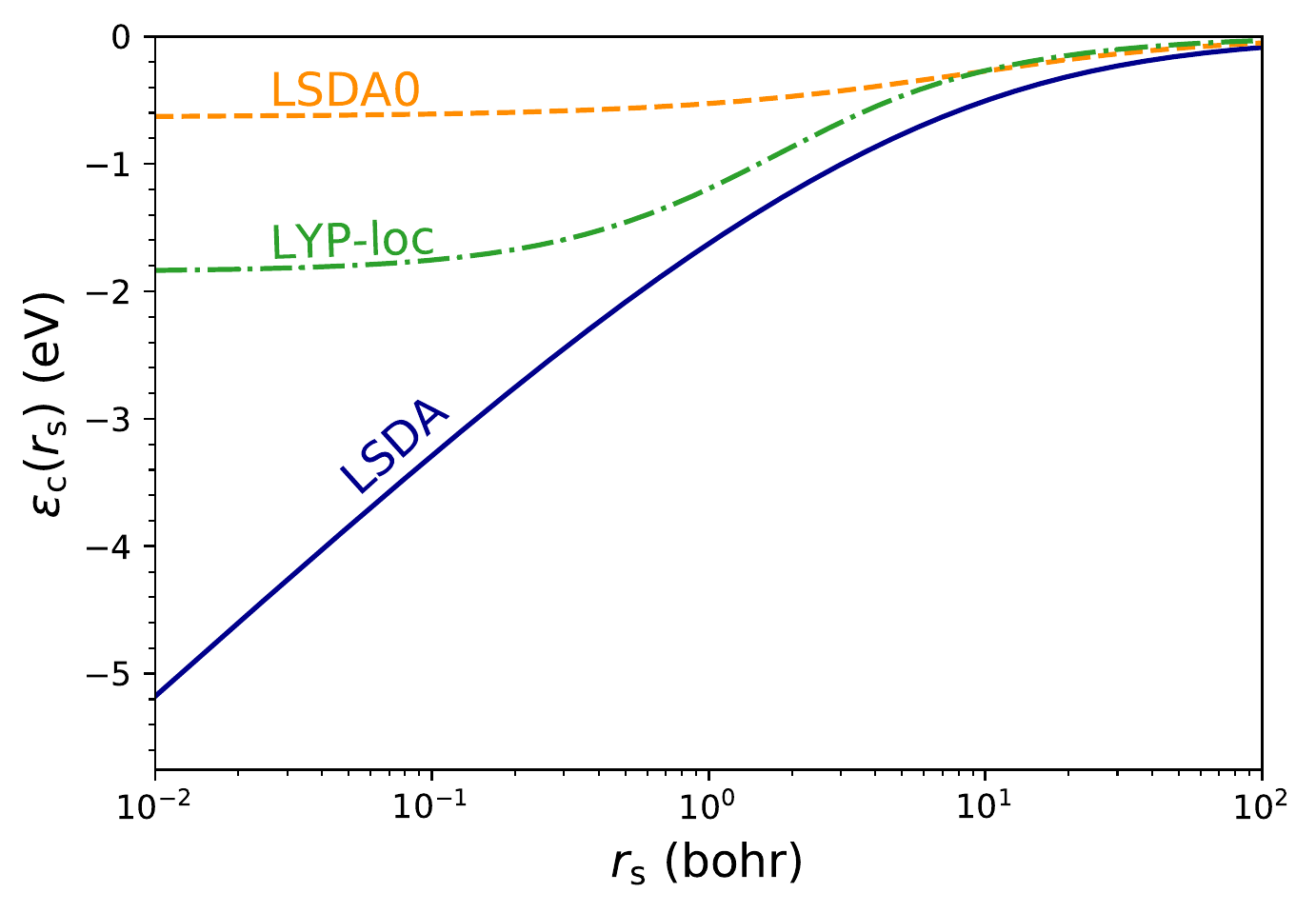}
  \caption{Comparison of the Perdew-Wang parameterization \cite{perdew1992} of the uniform electron gas LDA, the LYP LDA (LYP-loc) \cite{lee1988}, and the LSDA0 component of SCAN \cite{sun2016}.
  The LDAs for the correlation energy per electron $\varepsilon\suc$ (eV) are plotted as a function of the spin-unpolarized density $n=3/(4\pi \rs^3)$.
  }
  \label{fig:lda_comp}
\end{figure}

There are other ``LSDAs,'' as shown in Fig. \ref{fig:lda_comp}.
While the LSDA for the uniform electron gas is unimpeachable as the bedrock of solid state physics, it is a less obvious choice for organic chemistry.
The uniform-density limit of Lee-Yang-Parr correlation \cite{lee1988} (LYP-loc), and a form inspired by SCAN \cite{sun2016} (LSDA0) both provide more realistic descriptions of light atoms than the UEG LSDA.
Figure \ref{fig:lda_comp} shows that all three LSDAs correctly make the correlation energy vanish like $1/\rs$ as $\rs \to \infty$, but they have starkly different $\rs \to 0$ limits.
The LSDA diverges logarithmically \cite{gell-mann1957} as $\rs \to 0$, whereas LYP-loc and LSDA0 tend to finite constants, -1.845459 eV and -0.635397 eV respectively.
This logarithmic divergence becomes asymptotically correct in the limit of a neutral atom with $Z \to \infty$, but the actual correlation energies of neutral atoms converge very slowly to this limit \cite{kaplan2020}, with respect to $Z$.
LYP-loc and LSDA0 tend to finite constants and describe light atoms well.

The GGA level describes two distinct paradigms: GGAs for finite systems (atoms, molecules, and small clusters), and GGAs for extended systems (solids and surfaces).
Most GGAs for quantum chemistry or chemical physics tend to be of the former variety, such as Becke exchange \cite{becke1988} and Lee-Yang-Parr correlation \cite{lee1988,miehlich1989} (BLYP).
The PBEsol \cite{perdew2008} GGA defines the other extreme, prioritizing accuracy for solids.
The general-purpose PW91 \cite{perdew1991,perdew1992a} and PBE \cite{perdew1996} GGAs offer competitive accuracy for most systems.

With the advent of high performance computing, DFAs are generally assessed with self-consistent benchmarks for large sets of densities.
How systems are selected to form a representative set, and how error metrics are computed to best reflect both the accuracy of a DFA and the uncertainties in the reference values, is beyond the scope of this work.
Moreover, test sets are typically updated (with new systems, reference values, etc.), so we will give examples of families of sets and cite their most recent revision.
The Gaussian-$n$ (G$n$) series of test sets \cite{curtiss2007} \textit{currently} assesses the atomization energies of molecules composed of elements in the first three rows of the periodic table.
The S22 set \cite{marshall2011} of interaction energies assesses a DFA's description of weakly-bound, first and second row molecular complexes vital to biochemistry.

One of the most comprehensive quantum chemical test sets is the GMTKN, which currently \cite{goerigk2017} comprises 55 individual test sets (including the G2-1 and S22 sets) for a total of 2,544 single-point calculations.
By design, the GMTKN55 is predisposed to thermochemistry: the average $Z=4.14 \pm 4.73$ for an element within the set (51.5\% first-row, 44.9\% second-row, 3.0\% third-row elements).
The GMTKN also includes the MB16-43 set of artificial molecules to assess if a DFA can describe systems beyond those guiding its construction.

When assessed on some individual subsets of the GMTKN, like the G2-1 atomization errors, empirical GGAs like BLYP (4.7 kcal/mol mean absolute error or MAE) perform markedly better than non-empirical ones like PBE (8.6 kcal/mol MAE) \cite{ernzerhof1999}.
A popular error metric for the entire GMKTN55 is the WTMAD-2 \cite{goerigk2017}.
The WTMAD-2 scores of BLYP and PBE are 21.1 kcal/mol and 13.83 kcal/mol, respectively \cite{goerigk2017}.
For the MB16-43 set of artificial molecules, PBE makes a 22.78 kcal/mol MAE, roughly 2.5 times smaller than BLYP's 58.30 kcal/mol MAE \cite{goerigk2017}.
Thus non-empiricism lends itself to a better general description of many-electron systems.

\begin{table}
  \centering
  \begin{tabular}{lrrr} \hline
    meta-GGA & WTMAD-2 (kcal/mol) & With dispersion correction & Dispersion correction \\ \hline
    M06-L & 8.67 & 8.61 & D3(0) \\
    SCAN & 8.72 & 7.86 & D3(BJ) \\
    \rrscan & 8.8 & 7.50 & D4 \\ \hline
  \end{tabular}
  \caption{Comparison of a weighted error metric, WTMAD-2, for the entire GMTKN55 test set \cite{goerigk2017} for three meta-GGAs: M06-L \cite{zhao2006}, SCAN \cite{sun2015}, and \rrscan \cite{furness2020}.
  The type of dispersion correction is listed in the rightmost column.
  Data for M06-L and SCAN are taken from Ref. \cite{goerigk2017}, and data for \rrscan from Ref. \cite{ehlert2021}.
  }
  \label{tab:mgga_gmtkn}
\end{table}

Empirical meta-GGAs like M06-L have long been the most accurate meta-GGAs tested on the GMKTN; many are fitted to its subsets.
Table \ref{tab:mgga_gmtkn} shows that SCAN and \rrscan compete with M06-L for overall accuracy, and exceed it when using a (necessarily empirical) dispersion correction.
The dispersion-corrected \rrscan in particular competes with hybrid DFAs for overall accuracy.

A variant of \rrscan with a dispersion correction and basis-set corrections, called \rrscan{}-3c \cite{grimme2021} (WTMAD-2 of 7.5 kcal/mol), has the second-lowest WTMAD-2 for a SL DFA tested on the GMKTN55.
The highly-empirical B97M-V meta-GGA \cite{mardirossian2015} has the lowest overall SL WTMAD-2: 5.5 kcal/mol.
\rrscan{}-3c excels in predicting novel chemistry, with a low MAE of 12.1 kcal/mol for the the MB16-43 set, compared to B97M-V's 39.68 kcal/mol MAE \cite{najibi2020}.

This interpretation is borne out in simulations of liquid water.
SCAN is among the few DFAs to accurately predict the radial distribution functions of liquid water, and to make the I$h$ phase of ice less dense than liquid water \cite{chen2017}.
PBE, BLYP, and PBE0 all predict that ice is denser than liquid water.

Tests of solid state properties typically favor non-empirical DFAs.
There are fewer standard test sets for solid state physics, however we have advocated for the LC20 set of cubic lattice constants, bulk moduli, and cohesive energies \cite{sun2011}.
As no reliable theoretical reference values for solid-state properties exist, reference values should be taken from experimental data and corrected for zero-point vibrational motion of the nuclei, which are not typically included in practical DFT calculations.
Older tests, such as those of Ref. \cite{kurth1999}, which show that BLYP predicts equilibrium unit cell volumes with twice the error of PBE, are unlikely to have qualitative conclusions change when including zero-point effects, but the precise numerical values will change.
The average GGA errors in solid-state geometry are generally larger than the uncertainty in the reference values.
Modern meta-GGAs predict solid state lattice constants with sufficient accuracy that inclusion of zero-point effects can also change qualitative conclusions \cite{furness2020a}.
Modern tests \cite{sun2015,furness2020,furness2022} of solid-state geometries show that meta-GGAs like SCAN and \rrscan predict lattice constants with much greater accuracy than their empirical counterparts, like M06-L.

Recent work \cite{isaacs2018,kingsbury2022} has attempted to standardize large sets of solid-state properties, such as geometries, enthalpies of formation, magnetic moments, and band gaps.
This work has demonstrated that inclusion of exact constraints in the right manner is critical: \rrscan improves \cite{kingsbury2022} upon the PBE, PBEsol, and SCAN formation enthalpies despite satisfying one fewer constraint than SCAN.

The inclusion of exact constraints is critical for describing strongly-correlated materials, such as the high-$T_\mathrm{c}$ cuprate superconductors and transition metal oxides.
Few DFAs are able to recognize the undoped La$_2$CuO$_4$ as an antiferromagnetic insulator, and its Sr-doped counterpart as metallic, however SCAN does so without a need for empirical corrections \cite{furness2018}.
SCAN also provides a realistic description of the magnetic stripe patterns \cite{zhang2020} that emerge under Ba-doping.
SCAN provides a better starting point for studying structural and magnetic properties of transition metal oxides than the PBE GGA, but inclusion of a Hubbard-like $U$ correction brings SCAN's predictions close to experimental accuracy \cite{gautam2018}.
Moreover, the value of the $U$ correction is roughly 60\% that needed for the corresponding PBE+$U$ correction, suggesting SCAN much more accurately describes the electron-electron interactions in these complex materials than does PBE.

Fundamental bandgaps computed with SCAN and \rrscan are comparable and generally underestimate the experimental gap \cite{isaacs2018,kingsbury2022}.
The TASK \cite{aschebrock2019} exchange meta-GGA was constructed to accurately predict bandgaps in three-dimensional solids, but is paired with LSDA correlation, thus satisfying many fewer constraints than SCAN (although TASK satisfies all possible exchange constraints).
Thus one needs to select both a useful set of exact constraints and a mathematical construction that does so in a physical way when building DFAs.

Hybrid DFAs are the next rung above meta-GGAs, however in practice, there are fewer systematic improvements from the \textit{current} meta-GGA level to hybrids as there are from GGAs to meta-GGAs.
Recent meta-GGAs like \rrscan offer accuracy comparable to global hybrids at a small fraction of their computational cost.
The inclusion of exact exchange in a global hybrid reduces the self-interaction error of a semilocal DFA, and can therefore compensate, in part, for semilocal density-driven errors.

This reduction in self-interaction and density-driven errors is important \cite{janesko2008} for describing reaction barrier heights.
On the standard BH76 set \cite{peverati2011} of reaction barrier heights, self-consistent LSDA (PBE) makes a 12.7 kcal/mol (8.6 kcal/mol) MAE.
This is reduced to 5.5 kcal/mol (4.0 kcal/mol) by evaluating LSDA (PBE) on the self-consistent Hartree-Fock density \cite{janesko2008}.
Using updated BH76 reference data, PBE was found to make a 9.06 kcal/mol MAE, whereas its simplest global hybrid extension, PBE0 \cite{adamo1999}, only makes a 3.99 kcal/mol MAE \cite{wang2020}.

Density-driven errors (whether they arise from self-interaction errors or other ``delocalization'' errors \cite{li2017}) tend to dominate the errors of DFAs in hydrogen-bonded networks.
For example, applying a ``density-correction'' to SCAN (in this case, evaluating SCAN on the Hartree-Fock density) reduces SCAN's MAE in the binding energies of water clusters from a few kcal/mol to less than 1 kcal/mol \cite{dasgupta2021}.
The empirical $\omega$B97M-V (a range-separated hybrid based on the B97M-V meta-GGA) has the lowest WTMAD-2 on the GMKTN55 of any hybrid at 3.53 kcal/mol, but makes a 35.02 kcal/mol MAE on the MB16-43 set \cite{najibi2020}, much higher than the 12.1 kcal/mol \rrscan{} MAE.

Hybrids intended for solids, like HSE06 \cite{heyd2003}, tend to perform similarly to meta-GGAs and GGAs for lattice geometries.
In a test \cite{schimka2011} of the lattice constants of 30 common cubic solids (with zero-point anharmonic corrections to experimental values), HSE06 (MAE 0.033 \AA{}) was outperformed by PBEsol \cite{perdew2008} (MAE 0.022 \AA{}).
However, hybrids halve the typical GGA-level bandgap errors \cite{schimka2011,henderson2011}.

Hybrids can, however, exacerbate pathological behaviors of meta-GGAs, as observed for itinerant electron ferromagnets like Fe and Ni.
SCAN quite accurately predicts the geometries of these solids, but predicts magnetic moments that are too large compared to experiment \cite{ekholm2018,fu2018,mrt2019}.
Hybrids further worsen this trend, by predicting even larger magnetic moments \cite{fu2019}.

As of the time of writing, there are too few local hybrids, such as DM21 \cite{kirkpatrick2021}, and too few tests of them to compare with other DFAs.

There are rungs in Jacob's ladder above the hybrid level utilizing partial exact correlation from the unoccupied Kohn-Sham orbitals.
Prototypical examples include the random phase approximation (RPA) \cite{langreth1975} and G\"orling-Levy perturbation theory (GLPT) \cite{gorling1994}.
The RPA does not generally improve upon the GGA description of molecular atomization energies \cite{furche2001,schimka2013}, but does slightly improve upon the GGA level description of solid state lattice constants \cite{schimka2013}, albeit at much greater computational cost.
Grimme \cite{grimme2006} proposed including an approximation of the GLPT correlation energy to a hybrid DFA, yielding a ``double-hybrid,'' B2-PLYP.
More recent (empirical) double hybrids have been shown to have roughly 2 kcal/mol WTMAD-2 on the GMKTN55 \cite{goerigk2017}.
The RPA is the only partial-exact correlation method to have been applied to bulk solids.
The meta-GGA and hybrid DFA rungs of the ladder appear to be incomplete, unlike the GGA and LSDA rungs, thus we advocate for a better understanding of these lower-level DFAs before proceeding to the much more complex partial-exact-correlation methods.

There are also non-density functional methods that exactly solve the many-body Schr\"odinger equation, such as quantum Monte Carlo (QMC) \cite{ceperley1986}.
Many different QMC formulations exist, however all non-deterministically solve the Schr\"odinger equation.
QMC results have come into conflict with density-functional results for two simple systems: the interacting uniform electron gas (jellium) and the half-space jellium.
These are prototypical models of simple metals and their surfaces, respectively.
QMC is restricted to a finite number of electrons, and thus makes finite size errors when applied to (semi-)infinite systems like jellium and solids (or surfaces).

A recent QMC calculation \cite{spink2013} of infinite jellium found reasonable agreement with the long-standing Ceperley and Alder calculation \cite{ceperley1980} for densities $\rs > 2$, but deviated markedly \cite{bhattarai2018} from the exact $\rs\to 0$ limit known from many-body perturbation theory.
As no parameterization of the LSDA agrees with Ref. \cite{spink2013}, and the error in their calculation grows asymptotically as $|\ln \rs|$ as $\rs \to 0$ \cite{bhattarai2018}, it is much more likely that extant LSDA parameterizations are more accurate than the QMC calculation.
For the half-space jellium surface model, a QMC calculation combined release-node bulk energies with fixed-node finite-width slab energies \cite{acioli1996}, leading to surface formation energies much larger than those predicted by density functional methods \cite{constantin2008}.
A later QMC calculation \cite{wood2007} agreed well with the DFT surface energies.

Finally, we stress that tests on molecules alone or on solids alone do not provide a complete picture of an approximate functional.
Consider the long-range corrected hybrid functional LC-$\omega$PBE \cite{vydrov2006}, which 100\% of exact exchange at long range.
It has only one empirical parameter, the range $\omega$, and gives an excellent description of molecules, nearly satisfying \cite{vydrov2007} the fractional electron number constraint \textbf{n3} without machine learning.
Exact exchange at long-range is wrong for solids, especially metals, which are much better described by short-range--corrected hybrids like HSE06 \cite{heyd2003}, with 0\% exact exchange at long range.

Early density functional approximations in physics were based only on appropriate norms and exact constraints.
Axel Becke, who brought DFT into the mainstream of chemistry, and hybrid functionals into DFT, introduced a limited empiricism that was later greatly expanded by other functional developers.
In a recent article \cite{becke2022}, Becke argues that the only parameters that should be fitted to molecular data are those few that can be independently estimated from theory alone.
This minimally-empirical approach is shown to outperform the accurate, but highly-empirical, hybrid $\omega$B97X-V on the GMTKN55.

\begin{acknowledgements}

ADK thanks Temple University for a Presidential Fellowship.
ML thanks the Julian Schwinger Foundation for support.
JPP thanks the National Science Foundation (DMR 1939528) and the Department of Energy (CTC DE-SC0018331) for support.
\end{acknowledgements}

\bibliographystyle{apsrev4-2}
\bibliography{exac_cons}

\end{document}